\documentclass[aps,prd,amsmath,amssymb,twocolumn,preprintnumbers,preprint,nofootinbib,a4paper,11pt]{revtex4-1}
\pdfoutput=1
\usepackage{graphicx}
\usepackage{url}
\usepackage[bookmarks, pagebackref=true]{hyperref}
\usepackage{color}
	\definecolor{rossoCP3}{cmyk}{0,.88,.77,.40}
		\definecolor{graa}{rgb}{0.8,0.8,0.8}
		\definecolor{blaa}{rgb}{0.2,0.2,0.6}
		\hypersetup{
			colorlinks, 
			bookmarksopen, 
			bookmarksnumbered,
			citecolor=blaa, 		
			linkcolor=rossoCP3,	
			urlcolor=rossoCP3,			
			}
\usepackage{bm}
\usepackage{bbm}
\usepackage{pxfonts}
\usepackage[margin=5pt, font=small,labelfont=bf,justification=raggedright]{caption}
\usepackage{subfig}
\usepackage{youngtab}
\usepackage{slashed}

\usepackage{fancyhdr}
\newcommand{\beq}{\begin{eqnarray}}
\newcommand{\eeq}{\end{eqnarray}}

\newcommand{\bmp}{\noindent\begin{minipage}{16cm}}
\newcommand{\emp}{\end{minipage}\vskip 7mm} 

\newcommand{\qt}{\widetilde{q}}
\newcommand{\Tr}{\text{Tr}}
\newcommand{\yf}{\tiny \yng(1)}

\begin{document}
\phantom{g}\vspace{2mm}
\title{ \LARGE  \color{rossoCP3} Light Dilaton at Fixed Points \\ and \\ Ultra Light Scale Super Yang Mills} \author{Oleg {\sc Antipin}$^{\color{rossoCP3}{\varheartsuit}}$}\email{antipin@cp3-origins.net} 
\author{Matin {\sc Mojaza}$^{\color{rossoCP3}{\varheartsuit}}$}\email{mojaza@cp3-origins.net} 
\author{Francesco {\sc Sannino}$^{\color{rossoCP3}{\varheartsuit}}$}
\email{sannino@cp3-origins.net} 
\affiliation{
{\mbox {$^{\color{rossoCP3}{\varheartsuit}}$
 { \large \rm \color{rossoCP3}CP}$^{\color{rossoCP3}3}${ \large \rm \color{rossoCP3}- Origins}}} 
{\mbox{University of Southern Denmark, Campusvej 55, DK-5230 Odense M, Denmark.}}
}
\begin{abstract}
 We investigate the infrared dynamics of a nonsupersymmetric SU(X) gauge theory featuring an adjoint fermion, $N_f$ Dirac flavors and an Higgs-like complex $N_f \times N_f$ scalar which is a gauge singlet. We first establish the existence of an infrared stable perturbative fixed point and  then investigate the spectrum near this point. We demonstrate that this theory features a light scalar degree of freedom to be identified with the dilaton and elucidate its physical properties. We compute the spectrum and demonstrate that at low energy the nonperturbative part of the spectrum of the theory is the one of pure supersymmetric Yang-Mills. We can therefore determine the exact nonperturbative fermion condensate and deduce relevant properties of the nonperturbative spectrum of the theory. We also show that the intrinsic scale of super Yang-Mills is exponentially smaller than the scale associated to the breaking of conformal and chiral symmetry of the theory. 
 \\[.1cm]
{\footnotesize  \it Preprint: CP$^3$-Origins-2011-24}
\end{abstract}
 \maketitle
\thispagestyle{fancy}

\section{Introduction}

Understanding strong dynamics constitutes a continuous challenge. One facet of strong dynamics which is attracting much interest is the phase diagram of strongly interacting nonsupersymmetric gauge theories as a function of the number of flavors, colors and matter representations. The investigation of the phase diagram for nonsupersymmetric gauge theories, without fundamental scalars, for any matter representation, and several gauge groups, started in \cite{Sannino:2004qp,Dietrich:2006cm,Sannino:2009aw} and it was further investigated in \cite{Ryttov:2007sr,Ryttov:2007cx,Sannino:2010ue,Pica:2010mt,Antipin:2009dz, Antipin:2009wr,Pica:2010xq,Ryttov:2010iz,Mojaza:2010cm,Kaplan:2009kr}.  Besides the possibility of infrared fixed points recently it has been discovered that nonsupersymmetric gauge theories with fermionic matter at large number of flavors develop ultraviolet fixed points \cite{Pica:2010xq}.   
Furthermore, as soon as it was discovered that theories with an extremely low number of matter flavors were able to lead to large distance conformality and potentially become excellent candidates of models of dynamical electroweak symmetry breaking, the question whether these theories could feature light composite scalars was addressed in \cite{Hong:2004td,Dietrich:2005jn,Sannino:2009za}. These papers rekindled the long debate on the existence of a light scalar compared to the intrinsic scale of the theory. Their phenomenological relevance resides on the identification of this state with the composite Higgs in models of dynamical electroweak symmetry breaking featuring near conformal dynamics or the inflaton in models of successful minimal composite conformal inflation \cite{Channuie:2011rq}.  

A variety of arguments has been used in favor of the existence of a light scalar, not associated to Goldstone bosons linked to the spontaneous breaking of global symmetries.  Furthermore, there has been used a number of methods ranging from the use of the saturation of the trace and axial anomaly \cite{Sannino:1999qe} to supersymmetry \cite{Hong:2004td,Sannino:2003xe,Veneziano:1982ah,Armoni:2003gp}, alternative large N limits \cite{Sannino:2007yp,Sannino:2009za} as well as gauge-gauge duality \cite{Sannino:2009qc,Sannino:2009me,Mojaza:2011rw,Antipin:2011ny}. Other interesting attempts to investigate the dynamics associated to light scalars in near conformal field theories appeared in the literature \cite{Vecchi:2010dd,Hashimoto:2010nw,Choi:2011fy,Kutasov:2011fr}.

The suggestions of potential light dilatons for theories with near conformal invariance resided mostly on nonperturbative analysis  and the use of critical behavior which should be backed up by either lattice or  exact analytical computations. Here we investigate the perturbative and nonperturbative dynamics of a particular gauge theory, similar to the one investigated in \cite{Antipin:2011ny}, which allows us to clearly identify the dilaton and determine its properties using perturbation theory and even to be able to determine exactly some nonperturbative quantities arising at low energies.
Specifically, the theory we investigate here is an $SU(X)$ gauge theory with $N_{f}$ Dirac massless flavors, one adjoint Weyl fermion and a complex scalar singlet with respect to the gauge interactions, but bifundamental with respect to the nonabelian $SU(N_{f})_{L} \times SU(N_{f})_{R}$ global symmetries. 

In Section \ref{critical}, as a partial motivation for our new model and investigation, we provide a critical review of an analysis similar to ours \cite{Grinstein:2011dq}, but for a different gauge theory. 
We introduce the model in Section \ref{FFPs}, where we also establish the existence of interacting infrared fixed points in perturbation theory. Then in Section \ref{dil}, we determine the perturbative spectrum of states due to the Coleman-Weinberg  (CW) phenomenon \cite{Coleman:1973jx,Yamagishi:1981qq}. Here we provide the dilaton mass and identify the relevant physical state coupled to the dilatation current. We determine the region of stability of the CW potential and its geometric interpretation in Section \ref{GP}. We then discover that at energies much lower than the vacuum expectation value of the Higgs, the low energy theory is pure ${\cal N}=1$ SYM in Section \ref{Spectrum}. Here, we also show that the intrinsic renormalization group invariant scale of low energy SYM is exponentially smaller than the Higgs vacuum expectation value. We determine the gluino condensate as a function of  this scale. 
 In the Appendix we provide the explicit computations for the CW potential and conclude in Section \ref{Conclusions}.


\section{A comment on the Large $N$ and $N_f$ limit of the Grinstein and Uttuyarat Model }
\label{critical}
An analysis similar to the one performed here but for an entirely different theory has been done by Grinstein and Uttuyarat (GU) \cite{Grinstein:2011dq}. 
{ This pioneering work has triggered our interest in further exploring  near conformal dynamics in perturbative regimes. To better appreciate the model we will investigate in the next sections we start by commenting the large $N$ and number of flavors limits of   
the GU microscopic theory:} 
\begin{align}\label{L}
\mathcal{L} = 
&-\frac{1}{2} \Tr F^{\mu \nu}F_{\mu \nu} + i\overline{\psi}_j  \slashed{D} \psi^j + i \overline{\chi}_j \slashed{D} \chi^j + \frac{1}{2} (\partial_\mu \phi_1)^2  \nonumber \\ 
&+ \frac{1}{2} (\partial_\mu \phi_2)^2 
- y_1 (\overline{\psi}\psi + \overline{\chi}\chi) \phi_1 - y_2 (\overline{\psi}\chi + \overline{\chi}\psi) \phi_2 
\nonumber \\
& - \frac{1}{4!} \lambda_1 \phi_1^4 - \frac{1}{4!} \lambda_2 \phi_2^4
- \frac{1}{4} \lambda_3 \phi_1^2\phi_2^2 ,
\end{align}
where the fermions $\psi, \chi$ transform according to the fundamental representation of the $SU(N)$ gauge group and carry flavor index $j = 1, \ldots, N_f/2$. The scalar fields $\phi_1, \phi_2$ are real scalars and are gauge and flavor singlets.
The investigation of Banks-Zaks infrared fixed points requires that use of perturbation theory. This is achieved by arranging the  value $N_f/N$ to be proportional to $(1-\epsilon)$ with $\epsilon$ a small expansion parameter around the asymptotically free boundary. However, to make the $\epsilon$ parameter arbitrary small and continuous requires a well-defined large $N$ limit of the theory. This is achieved by rescaling the couplings as follows:
\begin{align}
g \to \frac{g}{\sqrt{N}}, \quad \text{and} \quad y_i \to \frac{y_i}{\sqrt{N_f \cdot N}}, \quad i = 1,2
\end{align}
Using these opportunely rescaled couplings, the GU fixed point \cite{Grinstein:2011dq} reads
\begin{align}
{g^*}^2 =  (4 \pi)^2 \frac{2}{75}\ \epsilon, 
 \  \quad {y_i^*}^2 = \frac{3x }{11}{g^*}^2,  \ 
\lambda_j^*= \frac{18}{11N} {g^*}^2
 \end{align}
 with $j=1,2,3$ and $x = N_f/N$. It is evident, in these variables, that the
nontrivial infrared fixed point in the quartic couplings $\lambda_j$  is lost in the large $N$ limit.  In order to claim the existence of such a fixed point in $\lambda_j$ one therefore must keep $1/N$ corrections. However, the order of the limits become important when taking this route. In fact, consider first the exact large $N$ (and therefore large $N_f$) limit. The infrared fixed point does not lead to quartic self-interactions and therefore one cannot observe chiral symmetry breaking induced by loop corrections \cite{Grinstein:2011dq}.  

For $N$ large but finite, needed for having a nonzero $\lambda^\ast_j$, the counting in the loop expansion and the $1/N$ needs to be addressed carefully given that terms such as:
\begin{equation}
\frac{\epsilon}{N} \ , \quad {\rm and} \quad \epsilon^2
\end{equation}
can be comparable.
This issue arises in the GU model because the scalars do no carry flavor index.  In the model we are about to introduce this issue is resolved yielding a natural large $N$ and $N_f$ limit.

\section{The Theory and its Fixed Points}
\label{FFPs}

The gauge theory we investigate is,
\begin{align}
\mathcal{L} = \mathcal{L}_K(G_\mu, \lambda_m, q, \qt, H) + y_H q H  \qt + \text{h.c.} \nonumber\\
- u_1 \left (\Tr [H H^\dagger]\right)^2 - u_2 \Tr \left [ ( H H^\dagger)^2 \right]  ,
\end{align}
 with the field content reported in Table~\ref{FieldContent}. Here  $\mathcal{L}_K$ summarizes the kinetic terms of the canonically normalized fields%
 \footnote{Note that in
 Ref.~\cite{Schechter:1971qa,Paterson:1980fc}, 
 the complex scalar fields $H_{ij}$ were not canonically normalized,
 why some results will deviate from those papers by numerical factors.}
  involving the covariant derivatives. 
  
\begin{figure}[tb]
\centering
\includegraphics[width=0.9\columnwidth]{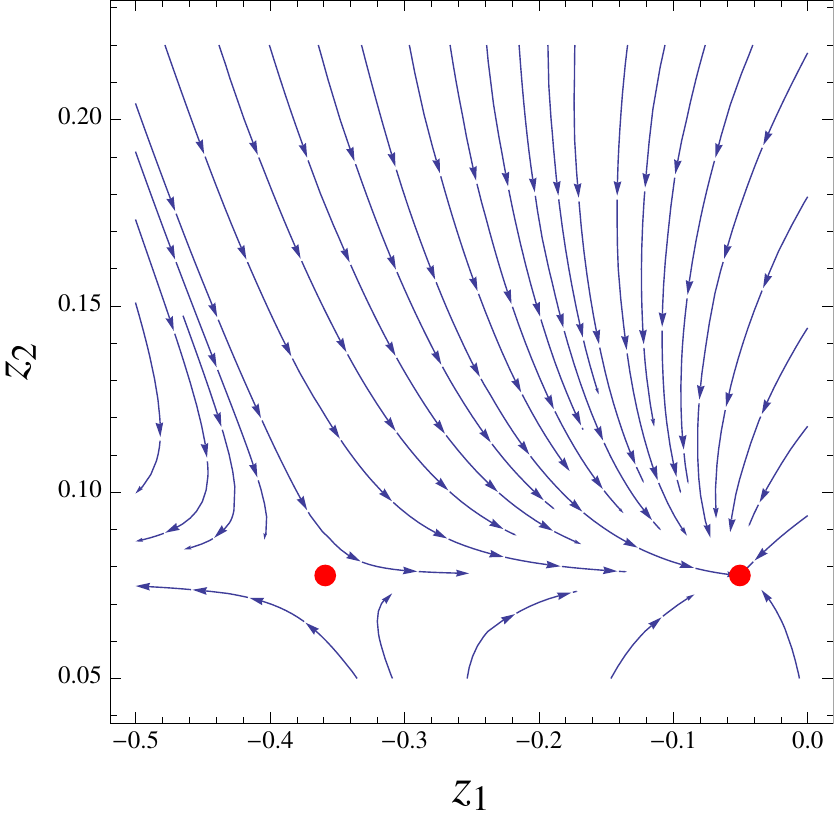} 
 \caption{RG flows for $\epsilon=0.1$ illustrating the (in)stability of the two fixed points. }
\label{RGdiagram}
\end{figure}

\begin{table}[b]
\vspace{-3mm}
\caption{Field content. The first three fields are Weyl spinors in the ($\frac{1}{2},0$) representation of the Lorentz group. $H$ is a complex scalar and $G_\mu$ are the gauge fields. $U(1)_{AF}$ is the extra \textbf{A}nomaly \textbf{F}ree symmetry
arising due to the presence of $\lambda_m$.}%
\vspace{-5mm}
\[ \begin{array}{c|c|c c c c} \hline \hline
{\rm Fields} &\left[ SU(X) \right] & SU(N_f)_L &SU(N_f)_R & U(1)_V& U(1)_{AF} \\ \hline 
\lambda_m & {\rm Adj} & 1 & 1 & 0 & 1 \\
 q &\yf &\overline{\yf }&1&~~\frac{N_f-X}{X} & - \frac{X}{N_f}  \\
\widetilde{q}& \overline{\yf}&1 &  {\yf}& -\frac{N_f-X}{X}& - \frac{X}{N_f}     \\
 \hline
  H & 1 & \yf & \overline{\yf} & 0 & \frac{2X}{N_f}\\
  G_\mu & \text{Adj} & 1 & 1 & 0 & 0 \\
   \hline \hline \end{array}%
\]%
\label{FieldContent}%
\vspace{-5mm}
\end{table}
  The gauge coupling constant  of the gauge group $SU(X)$ is identified with $g$.
The perturbative renormalization group equations (RGEs) of the coupling constants $g, y_H, u_1, u_2$
were derived in \cite{Antipin:2011ny, Chivukula:1992pm}.
Possible fixed points will be perturbative for $N_f$ near  $\frac{9}{2}X$, where the first coefficient of the gauge coupling beta function vanishes. It is therefore most convenient to fix $X$ and
define the small expansion parameter $\epsilon$ through $x \equiv N_f/X = \frac{9}{2}(1-\epsilon)$. As in any Banks-Zaks analysis $\epsilon$ and $x$ are considered continuous throughout the analysis. 
It is convenient to work with rescaled
coupling constants that
do not scale with $X$ and $N_f$. These read
\[
a_g = \frac{g^2X}{(4 \pi)^2} \ ,~\ a_H = \frac{y_H^2X}{(4 \pi)^2}\ ,~ z_1 = \frac{u_1N_f^2}{(4 \pi)^2}\  , ~z_2 = \frac{u_2N_f}{(4 \pi)^2} \ .
\]
Then dropping $1/X$-terms the perturbative beta functions 
are given purely in terms of the perturbative rescaled couplings and
the small parameter $\epsilon$ (through $x$):
\begin{subequations}
\label{betas}
\begin{align}
\beta(a_g) &= 
-2 a_g^2 \left[3-\frac{2 x}{3}+
\left(6-\frac{13 x}{3}\right)a_g +x^2 a_H
\right] \\[2mm]
\beta(a_H) &=   
 2 a_H \left[ \left(1 +x\right)a_H-3a_g\right] \\[2mm]
 \beta(z_1) &= 
 4(z_1^2 +4  z_1 z_2+3 z_2^2+z_1 a_H) \\[2mm]
 \beta(z_2) &= 
 4( 2 z_2^2+z_2 a_H-\frac{x}{2} a_H^2),
\end{align}
\end{subequations}
In this form, the beta functions are free of any explicit $X$ and $N_f$ dependency. This is an important feature which assures the smallness of $\epsilon$ to be arbitrary in all our results.
Notice that the quartic couplings do not contribute to the running of the gauge and Yukawa couplings to this level in perturbation theory.
To leading order in $\epsilon$, the system of RGEs has two real fixed points (FPs):
\begin{align}\label{FPs}
&a_g = \frac{11\epsilon}{9}, \quad a_H = \frac{2\epsilon}{3} \\
&z_1 = \frac{-2\sqrt{19}\pm \sqrt{2(8+3\sqrt{19})}}{6}\epsilon,  \quad z_2 = \frac{-1+\sqrt{19}}{6}\epsilon \nonumber
\end{align}
The FP corresponding to the upper sign for $z_1$, in the equation above, is all-directions (infrared) stable while the other FP has one unstable direction coinciding with the $z_1$-axis. This is illustrated in Fig.~\ref{RGdiagram} where we have kept gauge and Yukawa couplings at their FP values in Eq. \eqref{FPs} for $\epsilon=0.1 $. There is another solution with $z_2 =\frac{-1-\sqrt{19}}{6}\epsilon$ which however leads to complex values of $z_1$ and therefore is discarded. 
It is important to observe that the interacting FPs disappear if the adjoint fermion 
$\lambda_m$ is removed from the spectrum,
since then the coefficients of $\beta(a_g)$ will change. 
In particular, the perturbative regime would be moved to $x \approx \frac{11}{2}$
leading to a noninteracting  FP as will be shown explicitly in Section \ref{Spectrum}. 
 
\section{Lightest Scalar as the Dilaton}\label{dil}
In this section we will derive the scalar spectrum. We follow \cite{Gildener:1976ih} and summarize below the key points:

1) If the one-loop effective potential \mbox{$V_{\rm eff}=V_0+V_1$} can be calculated at some renormalization scale $M_0$ for which the tree-level term $V_0$ vanishes and is a minimum then one-loop perturbation theory can be used to show that $V_1$ causes $V_{\rm eff}$ to be negative and stationary. Therefore the global symmetries of the theory will be broken via the CW potential. 

2) The theory possesses, at least, one real light scalar corresponding to the field in the direction in scalar field space along which the potential develops the ground state.  It arises because $V_1$ breaks scale invariance of $V_0$ along this potential direction in field space. The mass of this state, to be identified with the \emph{dilaton}, is given by 
\begin{equation}
m_d^2=\frac{1}{8\pi^2 v^2} \sum_i \big[m_{scalar}^4 -4 m_{fermion}^4    \big]_i
\label{Dilatonformula}
\end{equation}
with $v$ as defined below\footnote{Quantum effects induce scalar masses of the order of the cut-off. Following the CW analysis, however we subtract these masses away \cite{Coleman:1973jx}. As  for \cite{Grinstein:2011dq} we are not solving the hierarchy problem.}. 

It was shown in Ref.~\cite{Paterson:1980fc} that in the \mbox{$U(N_f)\times U(N_f)$} linear sigma model, without fermions, the CW induced chiral symmetry breaking occurs when either condition A or B below applies:
\begin{itemize}
\item[A.] $z_2^0>0$ and $z_1^0+z_2^0=0$,
\item[B.] $z_2^0<0$ and $z_1^0+N_f z_2^0=0$,
\end{itemize}
where the superscript $^0$ refers to the quantities evaluated at the renormalization scale $M_0$.

In the large $X$ (and thus large $N_f$) limit only case A exists. We therefore restrict our analysis to case A.
In this case the vacuum expectation value (vev) of $H$ will be given by
\begin{align}
\langle 0|H_{ij}|0 \rangle=\frac{v}{\sqrt{2N_f}}\delta_{ij},
\label{vacuum}
\end{align}
with $v$ a real constant, while the factor $1/ \sqrt{2N_f}$ ensures the correct normalization of the field in the $\delta_{ij}$ direction. This expectation value breaks chiral symmetry to the diagonal subgroup, i.e. \mbox{$U(N_f)\times U(N_f)\to U(N_f)$}, leading to a multiplet of $N_f^2$ Goldstone bosons, a multiplet of $N_f^2-1$ heavy Higgses and 1 pseudo-Goldstone boson associated to the spontaneous breaking of scale invariance or dilatation symmetry, i.e. the dilaton with its mass given by Eq.~\eqref{Dilatonformula}.

After diagonalization of the scalar mass matrix, all $N_f^2-1$  heavy Higgses will have the same mass $m_H^2=2 u_2^0 v^2/N_f$ \cite{Eto:2009wu,Schechter:1971qa}.
The masses of the $N_f\times X$ copies of Dirac fermions, transforming according to the fundamental representation of $SU(X)$, is given by $m_f=y_H^0 v/\sqrt{2N_f}$. From Eq.~\eqref{Dilatonformula}, the mass of the dilaton reads (dropping $1/X$-terms for consistency):
\begin{equation}
m_d^2=\frac{32\pi^2 v^2}{N_f^2}\left [4 (z_2^0)^2 - x (a_H^0)^2   \right].
\label{Dilatonmass}
\end{equation}
{ We now provide, see the Appendix for further details, the scaling of $v$ with  $N_f$ \cite{Gildener:1976ih}.
The effective potential in the $\delta_{ij}$ direction at
the renormalization scale $M_0$ is  
\begin{align*}
V_{\rm eff}(\Phi \delta_{ij};M_0) = 
\frac{4 \pi^2}{N_f^2} &\Phi^4
\left[
4 z_2^2 \ln (4 z_2) - x a_H^2 \ln ( x a_H) \right . \nonumber \\
&\left . + (4z_2^2-x a_H^2) \ln  \frac{8 \pi^2 \Phi^2}{ N_f^2 \ M_0^2}
\right]_{M_0} \ \label{GWeffpot}
\end{align*}
where $\Phi$ is canonically normalized such that at the minimum
of the effective potential $\langle \Phi \rangle = v$. This value is
\begin{align}
v^2 =\frac{ M_0^2 N_f^2}{8 \pi^2} \  e^{\left \{  - \frac{1}{2} - \frac{4 z_2^2 \ln (4z_2) - x a_H^2 \ln (
 x a_H)}{4z_2^2-x a_H^2} \right \} } 
 \end{align}
Thus, all physical masses do not scale with $N_f$ 
demonstrating the large $X$ and $N_f$ consistency of the model.
Moreover, notice that at the point in parameter space where $m_d$
 vanishes, i.e. $4 z_2^2 = x a_H^2$, consistently also $v$ vanishes exponentially (it is the exponent to vanish not the prefactor). Henceforth, the model exhibits a phase transition between the chiral symmetric and chiral broken phases.}

We still need to show that the scalar with one-loop mass $m_d$ corresponds to 
the state which is created from the vacuum by the spontaneously broken dilatation current,
$D_\mu$ which is related to the trace anomaly of the improved energy-momentum tensor $\Theta_{\mu \nu}$ via
\begin{align}
\partial_\mu D^\mu = \Theta_\mu^\mu = \sum_c^{g, y_H, z_1, z_2} \beta(c) \frac{\partial \mathcal{L}}{\partial c} \ .
\end{align}
The right-hand side of this expression is easily computed 
and one finds, up to
terms that vanish via the equations of motion%
\footnote{In general, the trace anomaly involves wave function renormalization as well,
however, as noted in \cite{Grinstein:2011dq}, these terms vanish by using the equations
of motion.} 
\begin{align}
\Theta_\mu^\mu = &\frac{\beta(g)}{g} (F_{\mu \nu})^2 
+ \beta(y_H) (qH\widetilde{q} + \text{h.c.}) \nonumber \\
&- \beta(u_1) \left (\Tr [H H^\dagger] \right)^2
-\beta(u_2)\Tr \left [ ( H H^\dagger)^2 \right] 
\end{align}
At the level of perturbation theory we are working, 
the beta functions $\beta(g)$ and $\beta(y_H)$ 
are independent of the running of $u_1$ and $u_2$
 and in
particular drive  $u_1$ and $u_2$ towards the infrared stable FP.
We can therefore  neglect the first two terms in the above expression,
assuming $\beta(g)\approx 0$ and $\beta(y_H)\approx 0$ at the scale $M_0$.

The dilaton mass $m_D$ is defined  by the matrix element 
\begin{align}
\langle 0 \mid \Theta_\mu^\mu \mid D \rangle_{x=0} = - f_D m_D^2 \ ,
\label{Dilatonmass2}
\end{align}
where $f_D$ is the dilaton decay constant.
The pseudo-Goldstone boson of case A is parametrized by:
\begin{align*}
\phi_{ij} = \frac{\phi}{\sqrt{2 N_f}} \delta_{ij} ,
\end{align*}
with $\phi$  a real scalar field. 
Expanding $H$ on its mass eigenstates around the vacuum of Eq.~\eqref{vacuum} we have:
\begin{align*}
H_{ij} = \frac{v + \phi}{\sqrt{2 N_f}} \delta_{ij} + i\,\pi_{ij} + h_{ij},
\end{align*}
where $\pi_{ij}$ parametrizes the $N_f^2$ Goldstone bosons, and $h_{ij}$ parametrizes the $N_f^2-1$ massive eigenstates. It follows from orthogonality of this mass eigenbasis that $\pi_{ij}$ must be hermitian and $h_{ij}$ must be hermitian and traceless.
Then it is easy to check that only 
the field $\phi$ contributes linearly to the trace anomaly, i.e.
\begin{align}
\Theta_\mu^\mu = - \left(\frac{4 \pi}{N_f}\right)^2 &\left [\beta(z_1)  
+\beta(z_2) \right] v^3 \phi + \cdots
\label{traceanomaly}
\end{align}
where the ellipses stand for terms 
that contribute to the matrix element beyond tree-level.
From this, we realize that the dilaton state $| D \rangle$ can only be created by $\phi$ and this identifies $\phi$ as the dilaton. 

As a last non-trivial check we need to show that the dilaton mass $m_D$ in Eq.~\eqref{Dilatonmass2} matches the mass $m_d$ of $\phi$ in Eq.~\eqref{Dilatonmass}.
Taking the matrix element of Eq.~\eqref{traceanomaly} we find:
\begin{widetext}
\begin{align}
\langle 0 | \Theta_\mu^\mu | D \rangle \big |_{M_0} 
& = -\frac{64 \pi^2 v^3}{N_f^2}  \left ( z_1^2 + 5 z_2^2 + 4 z_1 z_2 - \frac{x}{2} a_H^2 \right ) \Big |_{M_0}
 = - \frac{32 \pi^2 v^3}{N_f^2} \left [ 4 (z_2^0)^2 - x (a_H^0)^2 \right] 
 = - v m_d^2  \equiv - f_D m_D^2, \nonumber \\
\end{align}
\end{widetext}
where it is clear that we can identify the mass of Eq.~\eqref{Dilatonmass} with the dilaton mass, i.e. $m_d^2 = m_D^2$, and $f_d = v$.                                                                                                                       

We discuss the remaining spectrum associated to the fermion in the adjoint representation, which remains massless, in a dedicated section where we use the power of supersymmetry to induce interesting features. 

\section{Geometrical Picture} \label{GP}

To gain more insight into the dilaton mass derived above we now turn to the geometrical interpretation of the CW  induced breaking of chiral symmetry using the method of Ref.~\cite{Yamagishi:1981qq}.  To the one-loop order the adjoint fermion $\lambda_m$ does not contribute to the effective potential, and we can adopt the results of Ref.~\cite{Chivukula:1992pm}, where the method of Ref.~\cite{Yamagishi:1981qq} was applied to the model without $\lambda_m$. The conditions for a minimum of the effective potential using the RG improved CW potential are (see also the Appendix):
\begin{subequations}\label{conditions}
\begin{align}
&z_1+z_2 < 0, \quad z_2>0 \label{c1}\\
&4(z_1+z_2)+ \beta(z_1)+\beta(z_2)=0 \label{c2}\\
&4[\beta(z_1)+\beta(z_2)]+\sum_{j}^{1,2} \sum_{c}^{z_1,z_2,a_H} \beta(c)\frac{\partial{\beta(z_j)}}{\partial c}>0 \label{c3}
\end{align}
\end{subequations}

\begin{figure*}[tb]
\centering
\includegraphics[width=0.6\textwidth]{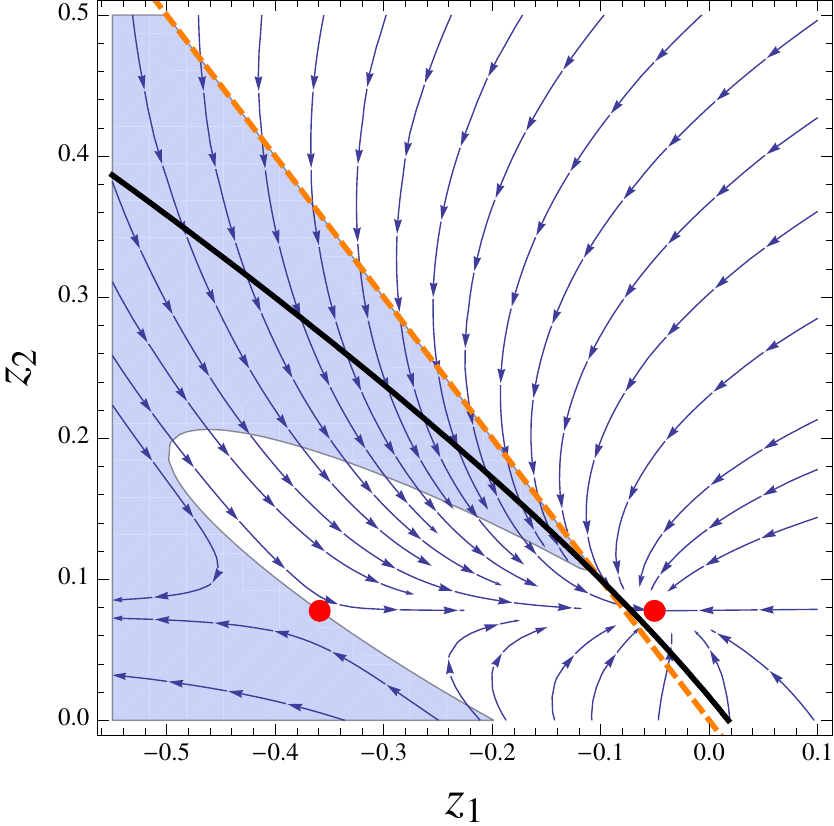} 
\caption{As in Fig.~\ref{RGdiagram}, but with the addition of the Coleman-Weinberg analysis. 
The thick (black) curve parametrizes the parameter values for which the effective potential has a minimum away from the origin. The latter is ensured to be a global minimum (within the shaded region) due to Eq. \eqref{conditions}: 
The shaded region is where simultaneously $V''(\langle H \rangle) >0$ 
and $V(\langle H \rangle) < V(0)$. The dashed (orange) line, defined by  $z_1 + z_2 =0$, marks where $V(\langle H \rangle) = V(0)$. The red RG trajectory, is the separatrix \cite{Braun:2011pp}, separating the phases of different infrared dynamics as explained in the text. The RG flow shown here is the one before taking into account spontaneous symmetry breaking which dramatically modifies it at scales below the spontaneous symmetry breaking scale $v$. The affected trajectories are the ones intersecting the black line.  
   }
\label{RGdiagram2}
\end{figure*}

Eq.~\eqref{c1} is a necessary condition for the minimum to be global,
i.e. $V(\langle H \rangle) < V(0)$, and
for chiral symmetry to be broken down to its diagonal subgroup, $SU(N_f)_V$.
Eq.~\eqref{c2} is the condition for $\langle H \rangle$ to be an extremum of the effective potential, i.e. $V_{\rm eff}'(\langle H \rangle) = 0$. We will refer to this condition as the \emph{stability line}. 
Eq.~\eqref{c3} is the condition for the extremum to be a minimum, i.e. $V_{\rm eff}''(\langle H\rangle) > 0$.
We argue that the new minimum exists as follows: The $\beta(c)\frac{\partial{\beta}}{\partial c}$ terms in Eq. \eqref{c3} are higher order in the couplings and therefore negligible to leading order in $\epsilon$.
Then the above conditions just tell us that the  allowed region of the stability line lies in the intersection of \mbox{$z_1+z_2 < 0$} and \mbox{$\beta(z_1)+\beta(z_2)>0$} in the coupling constant space.  
The situation is illustrated in Fig.~\ref{RGdiagram2} where the intersection of \mbox{$z_1+z_2 < 0$} and \mbox{$\beta(z_1)+\beta(z_2)>0$} is shaded and only the portion of the stability line (black) which falls into this region is plotted. 

{ Note that the RG flow shown in Fig.~\ref{RGdiagram2} must be amended when spontaneous symmetry breaking occurs, since at energy scales below $v$ several degrees of freedom decouple. The flow we show here is the one obtained in perturbation theory\footnote{We thank B. Grinstein for comments on this point.}.}

 The RG flow between the FPs does not generate a stable vacuum.  This is so due to the fact that 
the flow runs parallel to the $z_1$-axis and therefore the stability line is not crossed inside the shaded region. In this case the effective potential is not bounded from below.
{ In Fig.~\ref{RGdiagram2} we kept gauge and Yukawa couplings at their FP. All three curves (thick black, elliptic and dashed orange)  intersect at the point:}
\begin{equation}
 \left [\beta(z_1)+\beta(z_2)\right ]\Big |_{z_1=-z_2}=4z_2^2-xa_H^2=0\ .
\label{dilaton}
\end{equation}
Remarkably the condition for the dilaton to be massless, i.e. $m_d^2=0$ from Eq.~\eqref{Dilatonmass}, coincides with the boundary of the stability line Eq.~\eqref{dilaton}.
 {This boundary, a black point in Fig.~\ref{RGdiagram2}, uniquely defines an RG trajectory (the red one in the figure) separating the phase space of the CW induced symmetry breaking from the infrared conformal dynamics.} This line is known in the condensed matter literature as the {\it separatrix} \cite{Braun:2011pp}. 

{ Moreover, we discovered that the condition for the massless dilaton coincides with the one ensuring that the stability line (black thick line) intercepts $z_1+z_2=0$, the dashed orange line.} In the Appendix we show, that the one-loop contribution to the effective potential is $V_1\propto 4z_2^2-xa_H^2$. { Therefore the condition $m_d^2=0$ is equivalent to the statement that one-loop quantum corrections to the effective potential vanish. This implies that classical scale-invariance persists even after quantum corrections are taken into account.
Since, by selecting an appropriate RG trajectory, we may pass through the shaded region of Fig.~\ref{RGdiagram2} arbitrary close to the crossing point, the dilaton can be made arbitrary light.}

 
%

\section {Very Light Scale ${\cal N}=1$ Super Yang-Mills} \label{Spectrum}

The spectrum of our theory contains massive quarks and scalars whose mass scale is set by the vev scale, $v$, a nearly massless dilaton with mass given by Eq.~\eqref{Dilatonmass} and a pure ${\cal N}=1$ SYM sector composed of the remaining massless fields, the adjoint fermion $\lambda_m$ (gluino) and the gauge fields $G_\mu$ (gluons).

As we discussed in Sec.~\ref{dil}, the $v$ scale is set by the renormalization scale at which the tree-level potential vanishes. In terms of Fig.~\ref{RGdiagram2}, this is the scale where the RG trajectory crosses the dashed orange line parametrized by $z_1+z_2=0$.

 \begin{figure}[hbt]
\centering
 \includegraphics[width=0.9\columnwidth]{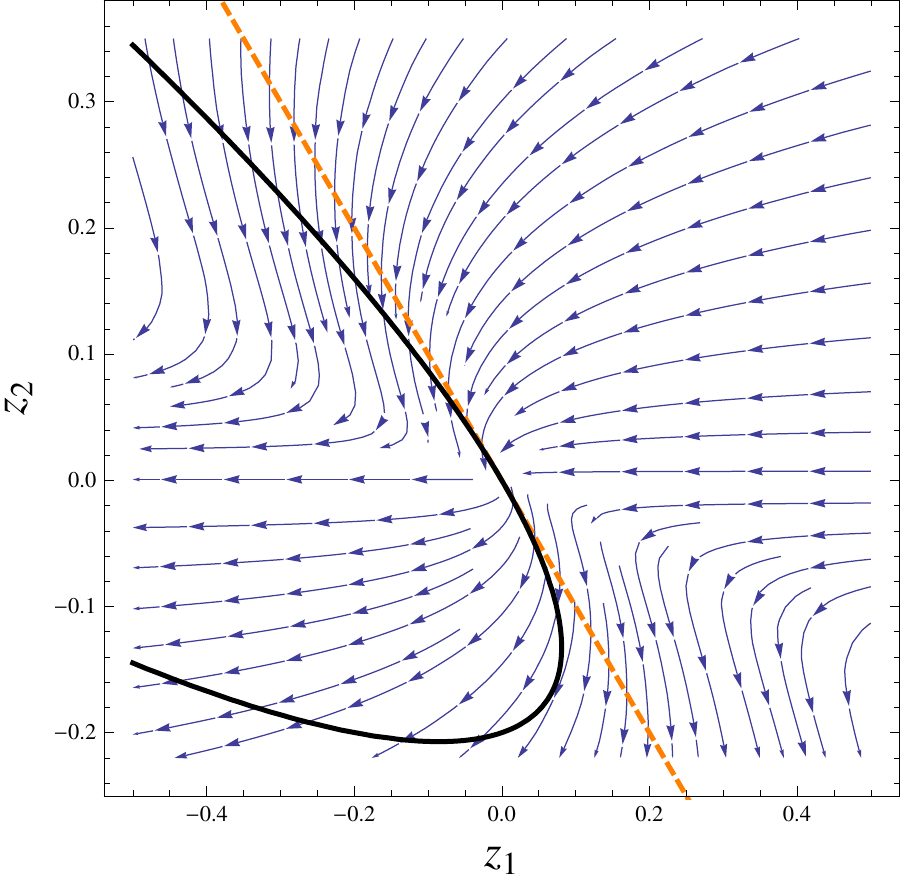} 
\caption{RG flows without Yukawa interactions leading to a non-interacting unstable fixed point at $(0,0)$. }
\label{RGdiagram1}
\vspace{-3mm}
\end{figure} 
The presence of the $\lambda_m$ state is crucial for the existence of an interacting infrared stable fixed point, despite the fact that it does not affect directly \eqref{Dilatonmass}.  In fact if $\lambda_m$ is removed from the spectrum of the underlying theory the one-loop perturbative fixed point in the Yukawa couplings in the remaining 
QCD-Higgs system disappears. To elucidate this point we show in Fig.~\ref{RGdiagram1} the RG flow for the $(z_1,z_2)$ system after having eliminated the Yukawa coupling. We observe only a non-interacting fixed point which coincides with the point where the dilaton is trivially massless. The adjoint fermion $\lambda_m$ is important to achieve a nontrivial dynamics. Furthermore  this model was also motivated by the recent gauge-gauge duality proposal    \cite{Mojaza:2011rw,Antipin:2011ny}.

We now turn to the low energy SYM nonperturbative properties of the theory. We can immediately determine exactly the renormalization group invariant scale of SYM indicated with $\Lambda_{\rm SYM}$ which is \cite{Novikov:1983uc}: 
\begin{eqnarray}
\Lambda^3_{SYM} & = & v^3 \left( \frac{16 \pi^2}{3X \, g^2(v)}\right) \, \exp \left(-\frac{8\pi^2}{X\,g^2(v)}\right) \nonumber \\
&=&v^3 \left( \frac{1}{3a_g (v)}\right) \, \exp \left(-\frac{1}{2a_g (v)}\right) \nonumber \\ & = &v^3 \left(\frac{3}{11 \epsilon} \right) \, \exp \left(-\frac{9}{22 \epsilon}\right) \ll v^3 \ .
\end{eqnarray}
Moreover: 
\begin{equation}
\langle \lambda_m \lambda_m \rangle =  - \frac{9}{32\pi^2} \Lambda^3_{SYM} \ .
 \end{equation}
 The low energy spectrum of SYM is constituted by a chiral superfield featuring a complex scalar (the gluino ball) and a Majorana fermion (the gluino-glue composite state) which because of supersymmetry are degenerate in mass. This sector is a non-perturbative one. The SYM low energy spectrum has masses proportional to $\Lambda_{SYM}$ which vanishes exponentially with $\epsilon$.  Because of supersymmetry there is no contribution to the trace of the energy momentum tensor. This is highly consistent with having assumed no contribution from the gluonic condensate to the trace of the energy momentum tensor. 

 \section {Conclusions}\label{Conclusions}
We introduced a computable model featuring a perturbative stable infrared attractive fixed point and determined the perturbative and nonperturbative spectrum near this fixed point.  We demonstrated that this theory features a light scalar degree of freedom with respect to the vacuum expectation value of the theory. We showed which state is to be identified with the dilaton and furthermore elucidated its physical properties. We also addressed the nonperturbative spectrum of the theory which is identified at low energy with the one of SYM. We finally determined the nonperturbative fermion condensate and extracted the relevant properties of the nonperturbative spectrum of the theory.   In the Appendix we provide useful definitions and explicit formulae to determine the effective one loop potential of the theory. Our findings demonstrate that there is a dilaton which is naturally identified with the Higgs of the theory, i.e. the field associated to the direction along which chiral symmetry breaks spontaneously. Remarkably the dilaton, in this calculable model, unless one fine-tunes the couplings at the point when it is exactly massless, is not the lightest state of the theory. The lightest states are the ones associated to the low energy SYM theory.

\appendix
\section{Effective Potential}
In this appendix we clarify the origin of some of the expressions used in the main text.

The Coleman-Weinberg (CW)  effective potential is defined by expanding
 the effective action around the \emph{classical field configuration} $\phi_c$, which in our 
case is:
\[
\phi_c^2 \equiv \frac{\Tr H_cH_c^\dagger}{N_f} \ .
\]
The effective potential obeys the renormalization group equation (RGE)
\begin{align}
\Big(M \frac{\partial}{\partial M}+\sum_{\lambda_i=u_1,u_2,y_H, g} {\beta} (\lambda_i)\frac{\partial}{\partial \lambda_i}+{\gamma}\phi_c \frac{\partial}{\partial \phi_c} \Big)V=0 ,
\end{align}
where the coefficients ${\beta}$ and ${\gamma}$ are 
respectively the beta function of the couplings and anomalous dimension of the scalar field $H$, i.e.
 \[
\frac{d\lambda_i}{d\ln M} = \beta(\lambda_i), \quad
\gamma = \frac{1}{2} \frac{d \ln Z_H}{d \ln M} \ , 
\]
with $Z_H$ the scalar wave function renormalization constant. Following Coleman and E. Weinberg \cite{Coleman:1973jx} it is convenient to use the dimensionless four point function: 
\[
V^{(4)} \equiv \frac{\partial^4 V}{\partial \phi_c^4} \ .
\]
For a generic $n$ point function the RGE is: 
\begin{equation}
\left(M \frac{\partial}{\partial M}   +  \sum_{\lambda_i=u_1,u_2,y_H, g} {\beta} (\lambda_i)\frac{\partial}{\partial \lambda_i}+n{\gamma}             \right) \Gamma^{(n)} = 0 \ , 
\end{equation}
which specialized to $V^{(4)}$ reads: 
\begin{equation}
\left(M \frac{\partial}{\partial M}   +  \sum_{\lambda_i=u_1,u_2,y_H, g} {\beta} (\lambda_i)\frac{\partial}{\partial \lambda_i}+4{\gamma}             \right) V^{(4)} = 0 \ .  
\end{equation}
Using the fact that $V^{(4)}$ is dimensionless it can only depend on the couplings and the dimensionless variable:
 \[
t = \ln \frac{\phi_c}{M} \ .
\]
Making use of the further relation
\begin{equation}
-(1 -\gamma) \, \partial t = \partial \ln M \ ,
\end{equation}
the RGE for $V^{(4)}$ reads:
\begin{align}
&\Big(-\frac{\partial}{\partial t}+\sum_{\lambda_i=u_1,u_2,y_H, g} \bar{\beta} (\lambda_i)\frac{\partial}{\partial \lambda_i}+4\bar{\gamma}\Big)V^{(4)}([\lambda_i],t)=0 ,
\label{RGEq} \\
&\text{where} \qquad \quad 
\bar{\beta} \equiv \frac{{\beta}}{1-{\gamma}}\ , \qquad \bar{\gamma} \equiv \frac{{\gamma}}{1-{\gamma}}\ . \nonumber
\end{align}
A useful  renormalization condition is the Coleman and E. Weinberg one:
\[
\lambda_i(0) = \lambda_i \ .
\]
Then the leading log solution of the RGE reads:
\begin{align}
 V_{\rm RG} = 
 &\left \{{u}_1([\lambda_i],t) (\Tr H H^\dagger)^2 \right . \nonumber\\
& \left . + u_2([\lambda_i],t) \Tr[H H^\dagger]^2 \right \} e^{4\int_0^t ds \bar{\gamma}([\lambda_i(s)]) } \ .
  \label{Veff}
\end{align}
We say that this is the renormalization group improved tree-level effective potential.
From this expression, it is now easy to derive the conditions in Eq.~\eqref{conditions} for a non-trivial minimum of the classical field away from the origin \cite{Yamagishi:1981qq,Bardeen:1993pj}. Note that this expression is consistent with a one-loop analysis of the running of the coupling constants $u_1$ and $u_2$.

To understand better the relation between the effective potential and 
the RG equations of the couplings and their relation to the dilaton mass, 
it is more appropriate to consider Gildener and S. Weinberg's approach to
the CW potential\cite{Gildener:1976ih}. 
 One chooses the renormalization scale $M$ in such a way that the tree level potential vanishes at this scale, i.e.
\begin{align}\label{GWconstraint}
V_{\rm RG} (M) = 0\ .
\end{align}
We are, therefore, expanding the potential perturbatively around the non-trivial vacuum of the classical field. We will be focusing on the case where $u_2 >0$. In Ref.~\cite{Paterson:1980fc} was shown that $H$ will have a minimum along its diagonal direction, i.e.
\[
\langle H_{ij} \rangle = v \delta_{ij}, \quad 
\phi_c = v.
\]
Then, it follows that the one-loop effective dimensionless potential evaluated  for this value of $H$ reads:
\begin{align}
&V^{(4)}([\lambda_i],t) =  V^{(4)}_0 +2 t  V^{(4)}_1   \nonumber\\
&V^{(4)}_0 = N_f (N_f u_1 + u_2) \label{1LEP}\\
&V^{(4)}_1 = \frac{1}{64\pi^2v^4} \sum_i [m_{scalar}^4 -4 m_{fermion}^4    ]_i  + \cdots \nonumber 
\end{align}
where the sum is over all mass eigenstates of scalars and Dirac fermions and the ellipses 
refers to scheme-dependent constant terms, which are not important for the following.
We can see how this potential is non-trivially related to the perturbative beta functions
in Eq.~\eqref{betas} by checking that it 
satisfies the RGE in Eq.~\eqref{RGEq}. 
As noted earlier, the mass eigenstates of the nondilatonic scalars and the Dirac fermions
are respectively $4u_2 v^2$ and $y_H v$ (note the difference in the normalization of $v$ from the main text). Thus, we find that
\begin{align}
\frac{1}{(4 \pi)^2}\frac{\partial V^{(4)}}{\partial t} 
&=\frac{ (N_f^2-1) (4u_2v^2)^2 - 4 N_f X (y_H v)^4}{2 (4 \pi)^4 v^4}  \nonumber\\
& =2 [ 4 z_2^2 - x a_H^2] - 8 \frac{z_2^2}{x^2X^2} \nonumber \\
 &\approx  2 [ 4 z_2^2 - x a_H^2], \label{RGE1}
\end{align}
where in the second line we have used the rescaled coupling constant and $x = N_f/X$ and in the third line for consistency dropped the $1/X$-term.
The last step is necessary, since to compute the other terms in the RGE we must use
the perturbative beta functions in Eq.~\eqref{betas}, while we note that the
one-loop expression of $\gamma = - a_H + \mathcal{O}(a_H^2)$.
Thus, to the order in perturbation theory we are working $\bar{\beta} = \beta $ and 
$\bar{\gamma} = \gamma$.
Then, the remaining terms of the RGE in terms of the rescaled couplings reads:
\begin{align}
&\Big (\sum_i \beta \frac{\partial}{\partial \lambda_i} + 4 \gamma \Big)\frac{V^{(4)}}{(4\pi)^4} = \nonumber \\
&=\beta(z_1) + \beta(z_2)
+ 4(z_1 + z_2)(-a_H) + \mathcal{O}(t) \nonumber \\
&= 4 \left ( z_1^2 + 5 z_2^2 + 4 z_1 z_2 - \frac{x}{2} a_H^2 \right ) \nonumber\\
& \to 2 [ 4 z_2^2 - x a_H^2], \label{RGE2}
\end{align}
where in the last line we have used the initial condition of Eq.~\eqref{GWconstraint} at
which we are evaluating the RGE,
which in this case is simply $z_1 = -z_2$.
We see that the first part of the RGE in Eq.~\eqref{RGE1} exactly cancels the second part
in Eq.~\eqref{RGE2}, as they should for Eq.~\eqref{1LEP} to be correct.
Notice that the one-loop coefficient $V_1$ of the effective potential is proportional
to the dilaton mass squared $m_d^2 \propto 4 z_2^2 - x a_H$, since from Eq.~\eqref{1LEP} and\eqref{RGE1} we find that:
\[
\frac{V_1^{(4)}}{(4\pi)^2} = 4 z_2^2 - x a_H^2 \ .
\]
 Thus, 
a vanishing dilaton mass corresponds to the vanishing of the one-loop correction
of the effective potential.

\end{document}